\documentclass[aps, prd, nofootinbib,superscriptaddress,preprintnumbers,letterpaper,natbib,nofootinbib,twocolumn]{revtex4-1}
\pdfoutput=1
\usepackage{amssymb,amsmath,latexsym,mathrsfs}
\usepackage{url}
\usepackage{enumitem}
\usepackage{graphicx}
\usepackage[usenames,dvipsnames]{xcolor}
\usepackage[breaklinks,colorlinks,urlcolor=blue,citecolor=blue,linkcolor=magenta]{hyperref}
\usepackage{multirow}
\usepackage{rotating}
\usepackage[caption=false]{subfig}
\usepackage{scalerel}
\usepackage[dvipsnames]{xcolor}
\usepackage{cleveref}
\usepackage{svg}

\newcommand{\Lagr}{\mathcal{L}}
\newcommand{\bino}{\tilde{B}}
\newcommand{\wino}{\tilde{W}}

\begin{document}
\title{Leptogenesis and Neutrino Masses Via Pseudo-Dirac Gauginos}

\author{Cem Murat Ayber }
\email{cemayber@cmail.carleton.ca}
\affiliation{Carleton University  1125 Colonel By Drive, Ottawa, Ontario K1S 5B6, Canada}

\author{Tammi Chowdhury}
\email{chowdh64@myumanitoba.ca}
\affiliation{University of Manitoba 66 Chancellors Cir, Winnipeg, Manitoba R3T 2N2, Canada}

\author{Seyda Ipek}
\email{Seyda.Ipek@carleton.ca}
\affiliation{Carleton University  1125 Colonel By Drive, Ottawa, Ontario K1S 5B6, Canada}

\begin{abstract}
 In a $U(1)_{R-L}$-symmetric supersymmetric model, pseudo-Dirac bino and wino can act like right-handed neutrinos, generating the light neutrino masses through a hybrid Type I + III inverse seesaw mechanism. We investigate such a model to accommodate the baryon asymmetry of the universe together with neutrino masses. A pseudo-Dirac gaugino goes under particle-antiparticle oscillations.  Possible $CP$ violation in bino decays, induced by mixing with the neutrinos, can be enhanced in bino--antibino oscillations. Focusing on a long-lived bino, we show that its oscillations and decays can generate the observed baryon asymmetry while the wino is responsible for generating the neutrino masses. This mechanism requires a decoupled mass spectrum with a bino of mass $M_{\tilde{B}}\sim \mathcal{O}({\rm TeV})$ and sfermions with mass $M_{\rm sf}\gtrsim 50$ TeV. Furthermore, for the bino to decay out-of-equilibrium before the electroweak sphalerons turn off, the messenger scale needs to be $\Lambda_M \sim \mathcal{O}(10^7~ {\rm TeV})$. We discuss the displaced vertex signals at the LHC resulting from such a high messenger scale.
\end{abstract}
\maketitle

\section{Introduction }

We observe a universe that consists of only matter and virtually no antimatter. This asymmetry is often quantified through the parameter $\eta$ given by \cite{ParticleDataGroup:2022pth,Planck:2018vyg} 
\begin{align}
    \eta=\frac{n_B-n_{\bar{B}}}{n_{\gamma}}=6.14\times10^{-10}\,,
\end{align}
where $n_B$, $n_{\bar{B}}$ are the baryon and antibaryon number densities respectively and $n_{\gamma}$ is the photon number density. Assuming symmetric initial conditions, in order to explain why there is now a baryon asymmetry, a particle physics model is required to satisfy the following ``Sakharov conditions" \cite{Sakharov:1967dj}. \textbf{1.} Baryon number violation, \textbf{2.} $C$ and $CP$ violation,
\textbf{3.} Out of equilibrium conditions.

Baryon number is anomalously violated in the Standard Model (SM) because the $SU(2)_L$ interactions are chiral. At temperatures  $130~{\rm GeV}< T < 10^{12}~{\rm GeV}$, $B+L$ number is violated through sphaleron processes, but at low temperatures these processes are exponentially suppressed~\cite{PhysRevLett.113.141602}. ($B-L$ is always conserved.)
$CP$ is also violated within the SM through an irreducible phase in the Cabibo-Kobayashi-Maskawa (CKM) matrix. Although this is one of the Sakharov conditions, the SM $CP$ violation generally is not sufficient to explain the baryon asymmetry of the universe (BAU) \cite{Gavela:1993ts, Huet:1994jb}\footnote{Recently, a BSM model that uses the observed $CP$ violation in the SM was introduced~\cite{Elor:2024cea}.}.
Finally, the SM does not satisfy the out-of-equilibrium conditions. Hence, beyond-the-Standard Model (BSM) physics is needed to explain the BAU.

In addition to failing to explain the BAU, the SM cannot explain the origin of neutrino masses either. The fact that neutrinos oscillate between their flavor eigenstates requires that they have masses.  Neutrino oscillation data give us two mass differences \cite{Esteban:2024eli}:
\begin{equation}
        \Delta m_{12}^2 \simeq 7.4\times 10^{-5}\ {\rm eV^2},\quad |\Delta m_{13}^2| \simeq 2.5\times 10^{-3}\ {\rm eV^2}.
    \label{eqn:numassdifferences}
\end{equation} 
Note that currently, the mass ordering and the absolute masses are not known, and the lightest neutrino is allowed to be massless. In order to explain how neutrinos acquire their masses, new physics (NP) is required. One of the most studied NP mechanisms to explain the origin of neutrino masses is the \emph{seesaw mechanism} with a heavy right-handed (RH) Majorana singlet fermion (see, e.g., \cite{KING:2016456, Gavela:2009cd, Xing:2021} and the references therein). In this scenario, the light neutrino masses are proportional to $\sim v^2/M$, where $v$ is the Higgs vacuum expectation value (vev) and $M$ is the Majorana mass of the RH neutrino. To match the mass scale of the SM neutrinos, the seesaw scale usually needs to be very high, $M\sim 10^{14}~{\rm GeV}$.
Alternatively, if RH neutrinos are pseudo-Dirac fermions, having both Dirac and Majorana masses, neutrino masses are generated via the inverse-seesaw (ISS) mechanism \cite{WYLER:1983205, Mohapatra:1986aw, Mohapatra:1986bd}. In ISS, light neutrino masses are instead proportional to a small Majorana mass that slightly breaks the lepton number.

A common approach to solving the origins of the BAU and the neutrino masses is \emph{leptogenesis}~\cite{Fukugita:1986hr}. (See \cite{Davidson_2008} for an extensive review.) In leptogenesis, a heavy RH Majorana neutrino decays out-of-equilibrium, producing a lepton asymmetry. The asymmetry in leptons is then partially converted into the baryon asymmetry through sphalerons. Minimal leptogenesis models require at least two RH neutrinos with masses $\mathcal{O}(10^{10-12}~{\rm GeV})$, which are impossible to probe in particle physics experiments. In models such as resonant leptogenesis~\cite{Pilaftsis:2003gt, Pilaftsis:2004xx, Pilaftsis:2005rv, Chauhan:2021xus, Klaric:2021cpi, King:2024idj, Das:2024gua} and ARS leptogenesis~\cite{Akhmedov:1998qx, Drewes:2017zyw, Drewes:2021nqr, Caputo:2018zky, Baumholzer:2018sfb}, the Majorana neutrino mass scale could be much lower, possibly $\mathcal{O}({\rm TeV})$. Generally, this lowering of the mass scale requires almost-degenerate RH neutrinos, resulting in an enhancement of the lepton asymmetry by the small mass difference between the RH neutrinos.

In this work, we analyze a leptogenesis scenario within a neutrino mass mechanism with pseudo-Dirac bino and wino~\cite{Coloma:2016vod, Ayber:2023diw}. Pseudo-Dirac gauginos are naturally expected in $U(1)_R-$symmetric minimally supersymmetric SM (MSSM), in which supersymmetric partners are charged under a global $U(1)_R$ symmetry. In \cite{Coloma:2016vod}, it was shown that a pseudo-Dirac bino can act like a (pseudo-Dirac) RH neutrino, generating light neutrino masses through an ISS mechanism. In \cite{Ayber:2023diw}, it was further shown that the pseudo-Dirac wino in the same model could also be involved in neutrino mass generation, giving rise to a hybrid type I+III ISS scenario with rich phenomenology. 

A pseudo-Dirac fermion can have particle--antiparticle oscillations, much like the neutral meson oscillations in the SM. Furthermore, if the oscillation frequency and the decay width of the particles are comparable, the effects of possible $CP$ violation in the system are enhanced~\cite{Ipek:2014moa}. In \cite{Ipek_2016}, such $CP$ violation via oscillations was shown to result in successful baryogenesis, given that the pseudo-Dirac fermion has $B$-violating decays. In the same work, a pseudo-Dirac bino in a generic $U(1)_R-$symmetric MSSM was discussed as a specific example, focusing on baryon number violation via $R-$parity violating interactions for a benchmark bino mass of $300$~GeV. 

In this work, we show that the hybrid bino-wino scenario in \cite{Ayber:2023diw} can lead to successful leptogenesis via bino--antibino oscillations in the early universe, explaining both the neutrino masses and the BAU. We explore the parameter space for successful leptogenesis, analyzing the bino mass between 500 GeV and 5 TeV, and the sfermion masses $\mathcal{O}(10-100~{\rm TeV})$.

The paper is organized as follows. In \Cref{sec:model} we describe the $U(1)_{R-L}$--symmetric SUSY model we work with. In \Cref{sec:oscillations}, we describe the bino--antibino oscillations in the early universe. In \Cref{sec:baryogenesis}, we calculate the baryon asymmetry and show our main results for the BAU in \cref{fig:basymlines,fig:deltaBvsm}. In \Cref{sec:signals}, we discuss the possible collider phenomenology resulting from this scenario. Finally, in \Cref{sec:conclusion} we give our conclusions and outlook.

\section{The Model}\label{sec:model}
In this section, we summarize the neutrino mass mechanism in a $U(1)_{R-L}$--symmetric SUSY model introduced in~\cite{Coloma:2016vod} and extended in~\cite{Ayber:2023diw}. For a viable leptogenesis scenario, we focus on~\cite{Ayber:2023diw}, which we will motivate shortly.

In MSSM with a global $U(1)_R$ symmetry, superpartners have $+1$ $R-$charges, while SM fields are not charged under $U(1)_R$. Due to this global charge, gauginos cannot have Majorana masses. In order to give mass to gauginos, adjoint Dirac partners with $-1$ $R-$charges are introduced. (See Table~\ref{tab:fields} for partial field content and charge assignments.) The relevant Dirac partners for this work will be the \emph{singlino} $S$ and \emph{tripletino} $T$, which are Dirac partners for bino and wino, respectively. 

\begin{table}
\centering
\begin{tabular}{|c|c|c|c|c|}
\hline
Superfields & $SU(2)_L$ & $U(1)_Y$	&	$U(1)_R$	&	$U(1)_{R-L}$ \\
\hline\hline
$L_{i}$	& 2& -1/2	&1	&	0 \\
$E_{i}^c$  &1 &1	&	1	&	2	\\	
\hline
$H_{u,d}$  & 2 & 1/2	&	0	&	0	\\
$R_{u,d}$	& 2 & -1/2 &	2	&	2	\\
\hline
$W_{\widetilde{B}}^\alpha $ & 1 & 0	&	1	&	1	\\
$\Phi_S$ &1 &0	&	0	&	0	\\
$W_{\widetilde{W}}^\alpha$ & 3 & 0	&	1	&	1	\\
$\Phi_T$ &3 &0	&	0	&	0	\\
\hline
\end{tabular}
\caption{{A partial superfield content of the model and their charge assignments. The $R-L$ charge is defined as the $U(1)_{R}$-charge minus the lepton number of the field. $L_i,\ E_{i}^{c}$ are the lepton superfields where the subindex \textit{i} indicates the fermion generation. The fermionic components of the superfields $R_{u,d}$ are the Dirac partners of the Higgsinos $H_{u,d}$. $\Phi_{S},\Phi_T$ are the superfields that have the same SM charges as $W^{\alpha}_{\tilde{B}}$ and $W^{\alpha}_{\widetilde{W}}$. Their fermionic components, $S$ and $T$, are the Dirac partners of the bino and the wino, respectively.}}\label{tab:fields}
\end{table}

Dirac gaugino masses can be generated via the soft terms~\cite{Fox:2002bu}
\begin{align}
    \int d^2\theta \frac{\sqrt{2}\,c_{\Tilde{\chi}}}{\Lambda_M} W'_\alpha W^\alpha_{\Tilde{\chi}} \Phi_{\Tilde{\chi}}\,, {\rm with~}M_{\Tilde{\chi}}\equiv\frac{c_{\tilde{\chi}}\,D}{\Lambda_M}\,,
\label{eq:Diracmass}
\end{align}
where $c_{\Tilde{\chi}}$ are dimensionless coefficients and ${\tilde{\chi}}=\tilde{B},\tilde{W},\tilde{g}$, for the bino, wino and gluino respectively. Here we assume that SUSY is broken in a hidden sector, via both $F-$ and $D-$terms, which communicate with the visible sector at a messenger scale $\Lambda_M$. $W'_\alpha =\theta_\alpha D$ is the field strength of a hidden sector $U(1)'$ that gets a $D-$term vev.

Although we started with a $U(1)$ symmetry, it must be broken due to gravity, like any other global symmetry. At the end, Majorana masses for gauginos will be generated via anomaly mediation~\cite{Randall:1998uk, Giudice:1998xp, Gherghetta:1999am},
\begin{align}
    m_{\tilde{\chi}}=\frac{\beta(g_{\tilde{\chi}})}{g_{\tilde{\chi}}}F_\phi\,, \label{eq:Majmass} \notag
\end{align}
where $\beta(g_{\tilde{\chi}})$ is the beta function of the relevant gaugino and $F_\phi$ is a conformal factor satisfying
\begin{align}
    \frac{m_{3/2}^3}{16\pi^2 M_{\rm Pl}^2}\lesssim |F_\phi|\lesssim m_{3/2}\,.
\end{align}
Here $m_{3/2}$ is the gravitino mass. Since the $U(1)_R$ is broken, we can also expect a Majorana mass for the Dirac partners,
\begin{align}
    \int d^2 \theta\, m_S \Phi_S^2~~~{\rm and}~~~\int d^2 \theta\, m_T \Phi_T^2\, .
\end{align}

In \cite{Ayber:2023diw}, the global symmetry was extended to a $U(1)_{R-L}$ symmetry, where $L$ is the lepton number. The resulting pseudo-Dirac electroweakino fields can act as right-handed neutrinos and generate non-zero neutrino masses when the following effective operators are considered:
\begin{align}
    \frac{d^{\, i}_{\eta}}{\Lambda_M}\int d^{2}\theta d^{2}\bar{\theta}\phi^\dagger \Phi_{\eta} H_u L_i , ~{\rm and}~ \frac{f^{\, i}_{\tilde{\chi}}}{\Lambda_M^2}\int d^2\theta\,  W'_\alpha W_{\tilde{\chi}}^\alpha H_u L_i, \notag 
\end{align}
where $\phi=1+\theta^2 F_\phi$ is the conformal compensator and $f^i_{\tilde{\chi}}, d^i_{\eta}$, for $\tilde{\chi}=\tilde{B},\tilde{W}, \eta=S,T$, and $i=e,\mu,\tau$ are dimensionless, complex coefficients. Note that in~\cite{Ayber:2023diw}, a scenario with $F_\phi = m_{3/2}$ was considered, which required $(m_{S}+m_T)\sim m_{3/2}\sim \mathcal{O}(100~{\rm keV})$ for $\Lambda_M\sim \mathcal{O}(500~{\rm TeV})$ to generate the correct neutrino mass scale. 
Later, we will see that for leptogenesis, we need a small mass splitting for the bino mass eigenstates, $m_{\tilde{B}} + m_S \sim F_\phi\sim (10^{-3}-10)$~eV, and a messenger scale $\Lambda_M\gtrsim 10^7$~TeV. Such a small $F_\phi$ with a large messenger scale would make neutrino masses far too small to match the neutrino oscillation data. Hence, we allow instead for the following operator to appear,
\begin{align}
    \frac{d^i_\eta \tilde{m}}{\Lambda_M}\int d^2\theta\,  \Phi_\eta H_u L_i\,,
\end{align}
where $\tilde{m} \gg F_\phi$ is some new mass scale. We leave its UV completion to future work. 

We can now write the part of the Lagrangian relevant for neutrino masses:
\begin{align}
    \Lagr\supset &\ \frac{1}{2}M_{\tilde{B}} \bino S + \frac12\left( m_{\bino} \bino \bino + m_S SS \right)  \notag\\ 
  &+\frac{1}{2}M_{\wino} \wino T + \frac12\left( m_{\wino} \wino \wino + m_T TT \right)  \notag\\ 
    &+  \frac{f^i_{\bino} M_{\bino}}{\Lambda_M} \bino h_u \ell_i +  \frac{d^i_S \tilde{m}}{\Lambda_M} S h_u \ell_i \notag \\
    &+  \frac{f^i_{\wino} M_{\wino}}{\Lambda_M} \wino h_u \ell_i +  \frac{d^i_T \tilde{m}}{\Lambda_M} T h_u \ell_i  +{\rm h.c.}
    \label{eq:massLag}
\end{align}
We ignore neutralino mixing and assume the two lightest gauginos are purely bino and wino. After EW symmetry breaking, this Lagrangian gives rise to the following neutrino-bino-wino mass matrix, in the $(\nu_{i},\bino,\wino, S, T)$ basis, 
\begin{equation}
\mathcal{M}_\nu= \begin{pmatrix}
\mathbf{0}_{3\times 3}		&	\mathbf{Y}_{\bino}v	&	\mathbf{Y}_{\wino} v	&	\mathbf{G}_{S}v 	& 	\mathbf{G}_{T}v \\
\mathbf{Y}_{\bino}^{T}v	&	m_{\bino}				&	0				 	&	M_{\bino}			&	0	\\
\mathbf{Y}_{\wino}^{T}v	&	0					&	m_{\wino}				&	 0					&	M_{\wino}	\\
\mathbf{G}_{S}^{T}v	&	M_{\bino}			&		0				&	m_{S}				&	0	\\
\mathbf{G}_{T}^{T}v	&	0					&	M_{\wino}			& 	0				 	& 	m_{T}	
\end{pmatrix},
\label{eqn:numassmatrix}
\end{equation}
where $Y^i_{\bino,\wino} = f^i_{\bino,\wino} M_{\bino, \wino}/\Lambda_M$ and $G^i_{S,T} = d^i_{S,T} \tilde{m}/\Lambda_M$. The mass matrix above has a hybrid type-I+III ISS structure, and in its most general form, it can generate all non-zero neutrino mass eigenvalues. In~\cite{Ayber:2023diw}, a simplified scenario with $\mathbf{G}_{S,T}=0$ was analyzed. The resulting light neutrino masses for normal ordering were determined to
be
\begin{equation}
\begin{aligned}        
m_1 = 0\,, \quad
m_{2,3}= \frac{v^2(m_S+m_T)}{\sqrt{2}\Lambda_{M}^2 } \sqrt{ 1 - 2\beta \mp \sqrt{1-4\beta}}\,,
\end{aligned}
\label{Eqn:numassNO} 
\end{equation}
where $m_2<m_3$. The mass-squared splitting ratios $\Delta m^2_{\rm solar}/\Delta m^2_{\rm atm}$, set the parameter $\beta \simeq 0.13$. Notably, the light neutrino masses are independent of the Dirac masses. 

In this work, focusing on generating the baryon asymmetry, we will allow all terms in the above mass matrix to be non-zero. This case is much more general than the simplified scenario described above, and an analytical solution for neutrino masses is not readily available. For the leptogenesis scenario that will be described shortly, we will consider a hierarchy where $m_S\ll \tilde{m} < m_T$. In this case, we expect the wino part of the mass Lagrangian in \cref{eq:massLag}, coming from the terms with $m_T$ and $G_T$, to be more important for generating the neutrino masses, while we will use the bino part for generating the baryon asymmetry. 

For successful leptogenesis, in order to have enough $CP$ violation, we will require $m_{\bino}+m_S\ll M_{\bino}$ and $G_S/Y_{\bino}\sim \tilde{m}/M_{\bino}\sim \mathcal{O}(10^{-4}-1)$. Furthermore, the out-of-equilibrium condition will be satisfied by the out-of-equilibrium decays of a long-lived bino. (Note that although the wino can be long-lived with a suitable choice of the particle spectrum, we focus only on the bino here.) Thus, an important quantity is the bino width. Assuming bino is the lightest neutralino, $M_{\bino}<M_{\tilde{W}}$, its width is
\begin{align}
    \Gamma_{\bino} \simeq \frac{M_{\bino}^3+M_{\bino}\tilde{m}^2}{8\pi\Lambda_M^2}\,. \label{eqn:binowidth}
\end{align}
The out-of-equilibrium condition can be met for $\Gamma_{\bino} < H(T=M_{\bino})$ where $H(T)$ is the Hubble expansion rate given by,
\begin{align}
    H(T)=\sqrt{\frac{4\pi^3g_*}{45}}\frac{T^2}{M_{\rm Pl}}~,
\end{align}
during radiation domination. Here, $g_\ast$ is the number of relativistic degrees of freedom, $\sim 100$ for the temperature scale we will consider, and $M_{\rm Pl}=1.2\times10^{19}~\rm{GeV}$ is the Planck mass. Additionally, we will require the bino to decay before the sphalerons turn off, $\Gamma_{\bino} < H(T=130~{\rm GeV}$.

\begin{figure}[t]
    \centering
    \includegraphics[width=\columnwidth]{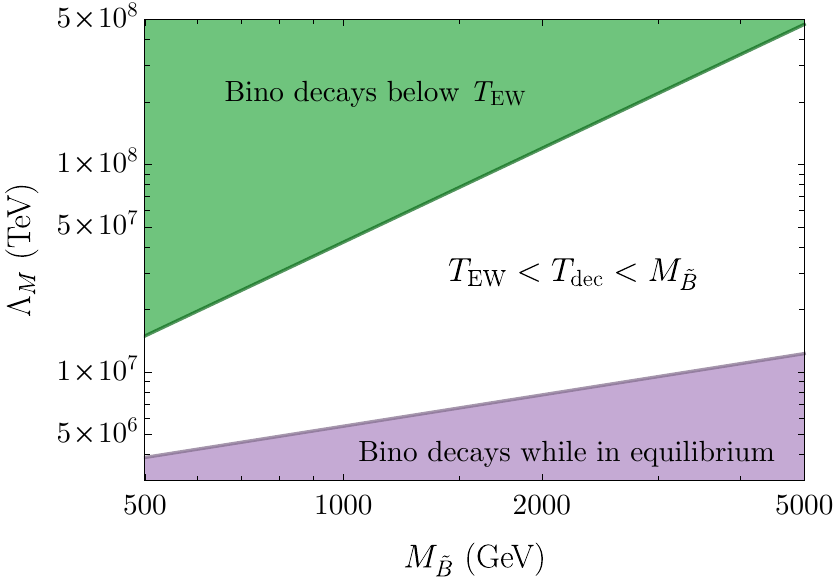}
    \caption{Regions of bino decay widths relevant for leptogenesis with respect to the bino mass $M_{\bino}$ and the messenger scale $\Lambda_M$. Here we take $\tilde{m}=0.2 M_{\bino}$.  In the bottom purple region, bino decays at a temperature above its mass, $T>M_{\bino}$. In the upper green region, bino lives long enough to decay below $T_{\rm EW}\simeq 130$~GeV, when sphalerons are no longer active. 
    }
    \label{fig:MBL}
\end{figure}

Thus, we consider a bino heavier than $500\rm~GeV$ and messenger scales $\Lambda_M \gtrsim \mathcal{O}(10^7~\rm TeV)$. (See Figure~\ref{fig:MBL}.) Furthermore, the amount of $CP$ violation needed to generate the correct baryon asymmetry will be selected for a (wide) range of $m_{\bino}+m_S$ and $\tilde{m}$. (See \cref{fig:moscillation}.) We also note that the condition on the lifetime is the reason why the pure-bino case in \cite{Coloma:2016vod} cannot account for both neutrino masses and the BAU. The parameters that allow for out-of-equilibrium bino decays result in a too small contribution to neutrino masses. In the bino-wino hybrid scenario of \cite{Ayber:2023diw}, there is enough freedom to accommodate both light neutrino masses and a long-lived bino. 
\section{Bino--Antibino Oscillations in the Early Universe} \label{sec:oscillations}

Pseudo-Dirac gaugino oscillations in vacuum at zero temperature were studied in the literature~\cite{Grossman:2012nn, Ipek:2014moa}. In~\cite{Ipek:2014moa}, it was shown that $CP$ violation in pseudo-Dirac gluino decays can lead to interesting same-sign dilepton signatures at the LHC. In the early universe, these oscillations would take place in a hot, dense medium in which the pseudo-Dirac binos interact efficiently. Furthermore, the expansion of the universe must be taken into account. Using the density matrix formalism~\cite{Tulin:2012re}, such an analysis was carried out in~\cite{Ipek_2016} for a baryogenesis scenario with $R$-parity violating bino interactions. (See also \cite{Chowdhury:2024lxx} for an expansion of that work.) Here we focus on a leptogenesis scenario in which the $L$-violating interactions also explain the origin of neutrino masses. In this section, we will briefly lay out the Boltzmann equations that govern the CP-violating pseudo-Dirac bino oscillations and decays in the early universe. 

\subsection{Bino oscillations}
 Before moving on to the early universe, let us summarize the oscillations of pseudo-Dirac binos, a phenomenon closely analogous to neutral-meson oscillations. We will be working with the Dirac field and its charge conjugate
\begin{align}
    \psi_{\tilde{B}} = \begin{pmatrix}
        \bino\\
        S^\dagger
    \end{pmatrix} \,, ~~~ \psi^C_{\tilde{B}} = C\overline{\psi}_{\tilde{B}}^T = \begin{pmatrix}
        S\\
        \bino^\dagger
    \end{pmatrix}\,,
\end{align}
which would be \emph{the} bino field if the $U(1)_{R-L}$ symmetry was exact. Note that the term ``bino" will be used both for $\psi_B$ and for its Weyl component $\tilde{B}$. This should be clear from the context. From the Lagrangian given in \cref{eq:massLag}, and ignoring the neutrino mass mixing, we can write the relevant part of the Hamiltonian in the $(\psi_{\bino},\psi_{\bino}^C)$ bases as
\begin{align}\label{eq:Ham}
    \mathbf{H}&=\mathbf{M}-\frac{i}{2}\mathbf{\Gamma}\,, \\
    &=    \begin{pmatrix}
    M_D&\hspace{-0.1cm} m\\
    m^*& \hspace{-0.1cm} M_D
    \end{pmatrix} - \frac{i}{2} \frac{M_{\bino}}{8\pi}
    \begin{pmatrix}
    |g_{\tilde{B}}|^2+|g_{S}|^2&\hspace{-0.2cm} 2g_{\tilde{B}}^*g_{S}\\
    2g_{\tilde{B}}g_{S}^*&\hspace{-0.2cm} |g_{\tilde{B}}|^2+|g_{S}|^2
    \end{pmatrix}, \notag
\end{align}
where $M_D\simeq M_{\bino}$ and $m\simeq(m_{\tilde{B}}+m_S^\ast)/2$ are the renormalized Dirac and Majorana masses\footnote{The Dirac mass is multiplicatively renormalized while the Majorana mass will get contributions from interactions. The loop contributions to the Majorana mass are $\delta\sim \frac{|g_{\bino}g_S^\ast|}{(4\pi)^2}M_D\sim \frac{M_{\bino}^3}{(4\pi)^2\Lambda_M^2}\lesssim \mathcal{O}(10^{-14}~{\rm eV})$.}, respectively, and $g_{\tilde{B}}= \sum_i f_i M_{\tilde{B}}/\Lambda_M$ and $g_S = \sum_i d_i\tilde{m}/\Lambda_M$. (We have taken the Dirac mass $M_{\bino}$ to be real.) As discussed before, we will consider binos that are heavier than $\sim 500$~GeV, and thus they can decay into final states with a Higgs boson, gauge bosons, and leptons via their mixing with the neutrinos. For the lepton asymmetry, we will only track the decays of $\tilde{B}, S \to h \ell$. 

Using the same notation as in the meson mixing case, we can easily find the \emph{heavy} and \emph{light} eigenstates of the above Hamiltonian,
\begin{align}
    |\psi_H\rangle=p|\psi_{\bino}\rangle-q|\psi_{\bino}^C\rangle , ~~ |\psi_L\rangle=p|\psi_{\bino}\rangle+q|\psi_{\bino}^{ C}\rangle\,,
\end{align}
where 
\begin{align}
   \left(\frac{q}{p}\right)^2=\frac{M_{12}^*-(i/2)\Gamma_{12}^*}{M_{12}-(i/2)\Gamma_{12}}~.
\end{align}
The corresponding eigenvalues are $\lambda_{H,L}=m_{H,L} -i\Gamma_{H,L}/2$. The mass and width differences between the heavy and light eigenstates are important for the oscillation dynamics and can be written as 
\begin{align}
    &\Delta m=\mathbb{R}(\lambda_H-\lambda_L), ~~~ \Delta \Gamma=-2\mathbb{I}(\lambda_H-\lambda_L)~, \\
&{\rm where}~~~
    \lambda_H-\lambda_L=2\sqrt{\left(M_{12}-\frac{i}{2}\Gamma_{12}\right)\left(M_{12}^*-\frac{i}{2}\Gamma_{12}^*\right)}~. \notag
\end{align}

Although both the mass and the width part of the Hamiltonian in \cref{eq:Ham} can be complex, there is only one physical phase in this system. We choose a real mass matrix, putting the $CP$ violating phase in $\mathbf{\Gamma}$. The decay matrix can be rewritten as
\begin{align}
    \mathbf{\Gamma}=\Gamma_{\bino} \begin{pmatrix}
        1&2r e^{i\phi}\\
        2r e^{-i\phi}&1
    \end{pmatrix}\,,
\end{align}
where $\Gamma_{\bino}$ is the bino decay width given in \cref{eqn:binowidth} and
\begin{align}
    r=\Big|\frac{g_{\bino}g_S}{g_{\bino}^2 + g_{S}^2}\Big| \simeq \frac{M_{\bino}\tilde{m}}{M_{\bino}^2+\tilde{m}^2}\,.
\end{align}

In the leptogenesis section, we will set $\tilde{m}\simeq0.2 M_{\bino}$, i.e. $r\simeq0.2$ as a benchmark value. This choice leads to a large $CP$ violation, but we will show that the observed BAU can also be generated with smaller values, $r\sim 10^{-4}$.

The eigenstates of the Hamiltonian are a superposition of bino and antibino states. A purely particle state at $t=0$ will evolve to acquire an antiparticle component, and vice versa for an initial antiparticle state. The time evolution of an initial particle or antiparticle state is given by
\begin{align}
\begin{split}
        |\psi_{\bino}(t)\rangle&=g_+(t)|\psi_{\bino}\rangle-\frac{q}{p}g_-(t)|\psi_{\bino}^C\rangle \,,\\
      |\psi_{\bino}^c(t)\rangle&=g_+(t)|\psi_{\bino}^C\rangle-\frac{p}{q}g_-(t)|\psi_{\bino}\rangle\,,
\end{split}
\end{align}
where 
\begin{align}
    g_{\pm}(t)=\frac{1}{2}\left(e^{-im_Ht-\frac{1}{2}\Gamma_Ht}\pm e^{-im_Lt-\frac{1}{2}\Gamma_Lt}\right)\,.
\end{align}

We quantify the relevant $CP$-violation in the decays of the binos by the following parameter;
\begin{align}
    \varepsilon = \frac{(N_\ell + \bar{N}_\ell) - (N_{\bar{\ell}} + \bar{N}_{\bar{\ell}}) }{N_\ell + \bar{N}_\ell + N_{\bar{\ell}} + \bar{N}_{\bar{\ell}}}\,, \label{eq:epsilon}
\end{align}
where 
\begin{align*}
    N_\ell = \int_0^\infty dt \frac{d\Gamma (\psi_{\bino} \to h\ell)}{dt}\,,~  \bar{N}_\ell = \int_0^\infty dt \frac{d\Gamma (\psi_{\bino}^C \to h\ell)}{dt}\,, 
\end{align*}
and $\Gamma (\psi_{\bino} \to h\ell)$ is the partial decay width for an initial $|\psi_{\bino}\rangle$ state to decay into a lepton state. Note that the above measure of asymmetry is time-integrated. Setting $\Delta\Gamma/(2\Gamma_{\bino}) =0$,  in the limit $r\ll 1$ and $m \gg r\Gamma_{\bino}$, we have the following approximation,
\begin{align}
    \varepsilon \simeq \frac{2\sin\phi \,x\,}{(1+x^2)}\frac{r(1-r^2)}{(1+r^2)}\,,\label{eqn:CPsimplified}
\end{align}
where $x\equiv \Delta m/\Gamma_{\bino}\simeq 2m /\Gamma_{\bino}$. As expected, the $CP$ violation is maximized for $x\sim 1$. However, the parameter regime we consider will be where $x\simeq \mathcal{O}(1-1000)$, which still results in enough baryon asymmetry. Since this approximation is valid in the parameter regime we consider, we will use \cref{eqn:CPsimplified} in the calculation of the baryon asymmetry later in \cref{eq:leptonBoltzman}.

\subsection{Other bino interactions}

Unlike generic RH neutrinos in various seesaw models, the bino in this model has copious interactions with the SM particles.  Elastic scatterings of the bino with the SM fermions in the plasma and bino--antibino annihilations affect the bino--antibino oscillations in the early universe, and hence the baryon asymmetry that will be generated through their decays. Assuming a common sfermion mass, $M_{\rm sf}$, the relevant effective interaction Lagrangian, after integrating out heavy sfermions, can be written as \cite{Hsieh:2007wq}, 
\begin{align}
   - \mathcal{L}_{\rm int} &= \frac{2g_Y^2}{M_{\rm sf}^2}\left( Y_L^2 \bar{\psi}_{\bino}^C P_L F \bar{F} P_R \psi_{\bino}^C+  Y_R^2 \bar{\psi}_{\bino} P_R F \bar{F} P_L \psi_{\bino}\right),
\end{align}
where $F^T = (f_L,~ \bar{f}^\dagger_R)$  and $Y_{L, R}$ are the hypercharges for the left- and right-handed SM fermions, $f_L$ and $f_R$, respectively. We can rewrite the above Lagrangian, using Fierz identities, to read, 
\begin{align}
     - \mathcal{L}_{\rm int} &= \frac{g_Y^2 Y_L^2}{M_{\rm sf}^2}\bar{\psi}^C_{\bino}\gamma_\mu P_L \psi_{\bino}^C\,\bar{F}\gamma^\mu P_R F\,, \notag\\  
   &\quad \quad+ \frac{g_Y^2 Y_R^2}{M_{\rm sf}^2}\bar{\psi}_{\bino}\gamma_\mu P_R \psi_{\bino}\bar{F}\gamma^\mu P_L F\,, \notag \\
   &= \frac{g_Y^2}{2M_{\rm sf}^2} \bar{\psi} \gamma_\mu P_R \psi\, \bar{F}\gamma^\mu (g_V -g_A \gamma_5)F\,, \label{eq:Lint}
\end{align}
where $g_V = Y_R^2 - Y_L^2$ and $g_A = Y_R^2 + Y_L^2$.

 For analyzing the bino--antibino oscillations in the early universe, it is useful to divide the above interaction Lagrangian into two parts, for \emph{flavor blind} and \emph{flavor sensitive} interactions,
 \begin{align}
 \begin{split}
          \mathcal{L}_{\rm int} &= \mathcal{L}_{\rm blind} + \mathcal{L}_{\rm sens}\,, \\
    \mathcal{L}_{\rm blind} &= \frac{g_Y^2}{4M_{\rm sf}^2}\bar{\psi}_{\bino}\gamma_\mu\gamma_5\psi_{\bino} \bar{F}\gamma^\mu (g_V-g_A\gamma_5)F\,, \\  
    \mathcal{L}_{\rm sens}  &= \frac{g_Y^2}{4M_{\rm sf}^2}\bar{\psi}_{\bino}\gamma_\mu\psi_{\bino} \bar{F}\gamma^\mu (g_V-g_A\gamma_5)F\,.  
\label{eq:Lint2}
  \end{split} 
 \end{align}
\noindent  $\mathcal{L}_{\rm blind}$ is symmetric under $\psi_{\bino} \to \psi_{\bino}^C$ and $\mathcal{L}_{\rm sens}$ is antisymmetric. (Note that $F$ is kept unchanged.) The thermally-averaged annihilation and scattering cross-sections for flavor-blind and flavor-sensitive interactions are 
\begin{align}\label{eq:xsections}
    \begin{split}
        \langle\sigma\,v\rangle_{\rm sens}^{\rm ann}&=  \frac{C\,M_{\bino}^2}{4M_{\rm sf}^4}\,, \quad
          \langle\sigma\,v\rangle_{\rm blind}^{\rm ann}=   \frac{C\,M_{\bino}^2}{4M_{\rm sf}^4}\frac{1}{z}\,, \\
        \langle\sigma\,v\rangle_{\rm sens}^{\rm scat}&=  \frac{C\,M_{\bino}^2}{4M_{\rm sf}^4}\frac{3z^2+6z+7}{6(z^4+z^3)} \simeq  \frac{C\,M_{\bino}^2}{4M_{\rm sf}^4} \frac{1}{2z^2}\,, \\
        \langle\sigma\,v\rangle_{\rm blind}^{\rm scat}&= \frac{C\, M_{\bino}^2}{4M_{\rm sf}^4}\frac{9z^2+6z+2}{6(z^4+z^3)} \simeq  \frac{C\,M_{\bino}^2}{4M_{\rm sf}^4} \frac{3}{2z^2}\,,
    \end{split} 
\end{align}
where 
\begin{align*}
    C = \frac{g_Y^4}{4\pi}\left(\sum_{i={\rm leptons}} \hspace{-0.1cm} \left(Y_{Ri}^4+Y_{Li}^4\right) + 3 \sum_{i={\rm quarks}} \hspace{-0.1cm}  \left(Y_{Ri}^4+Y_{Li}^4\right) \right).
\end{align*}
The sums run over all the quark and lepton flavors. We take all SM fermions, including the top quark, to be massless, since we consider $M_{\bino}\sim \mathcal{O}({\rm TeV})$. 

\subsection{Oscillations in the early universe}

The $CP$ violation necessary for generating the BAU requires that bino--antibino oscillations take place in the early universe. We follow the density matrix formalism described in \cite{Tulin:2012re} in order to incorporate the oscillation dynamics in the early universe. This method was developed for neutrino oscillations in the medium and the early universe~\cite{Dolgov:1980cq, Sigl:1993ctk} and was later applied to asymmetric dark matter models~\cite{Buckley:2011ye, Cirelli:2011ac, Tulin:2012re}. 

The following Boltzmann equation describes the evolution of the bino density matrix ${\rm \textbf{Y}}(\equiv n/s \propto \sum_{\psi_{\bino},\psi_{\bino}^C} f_{ij}|i\rangle \langle j|$ with the generalized quantum distribution functions $f_{ij}$) in the non-relativistic regime,
\begin{align}
    \begin{split}
    zH\frac{d\mathbf{Y}}{dz}=-i(\mathbf{HY}-\mathbf{YH^\dagger})-\sum_{+,-}\frac{\Gamma_\pm}{2}\left[O_\pm\left[O_\pm,\mathbf{Y}\right]\right]\\-\sum_{+,-}s\langle\sigma v\rangle_\pm\left(\frac{1}{2}\{\mathbf{Y},O_\pm\mathbf{\bar{Y}}O_\pm\}-Y_{\rm eq}^2\right), 
    \end{split} \label{eq:boltzmann}
\end{align}
where $z=M_{\bino}/T, s=2\pi^2g_*T^3/45$ is the entropy density, $\mathbf{H}$ is the Hamiltonian given in \cref{eq:Ham} and
\begin{align}
    Y_{\rm eq} = \frac{n_{\rm eq}}{s} = \frac{45}{4\pi^3\sqrt{2\pi}g_\ast}z^{3/2}e^{-z},
\end{align}
is the equilibrium abundance for $z>1$. $O_\pm = {\rm diag}(1,\pm 1)$ and the subscript $\pm$ differentiate between interaction terms that are symmetric $(+)$ or antisymmetic $(-)$ under the transformation $\psi_{\bino}\to \psi_{\bino}^C$. For binos, both types of interactions are present. (See \cref{eq:Lint2}.) Note that the above equation is only valid for non-relativistic particles, ignoring effects $\mathcal{O}(p^2/M^2)$, where $p\sim T$. (The Hamiltonian is given for $\mathbf{p}=0$.) Hence, we only use this equation for $z=M_{\bino}/T> 1$. The density matrices are
\begin{align*}
    \mathbf{Y} =\begin{pmatrix}
        Y_{\psi_{\bino}} & Y_{\psi_{\bino} \psi_{\bino}^C}\\
        Y_{\psi_{\bino}^C \psi_{\bino}} & Y_{\psi_{\bino}^C}
    \end{pmatrix}\,,~~\mathbf{\bar{Y}} =\begin{pmatrix}
        Y_{\psi_{\bino}^C} & Y_{\psi_{\bino}^C \psi_{\bino}}\\
        Y_{\psi_{\bino} \psi_{\bino}^C} & Y_{\psi_{\bino} }
    \end{pmatrix}\,,
\end{align*}
whose diagonal elements correspond to the bino and antibino abundances, e.g. $Y_\psi = n_{\psi_{\bino}}/s$. 

In \cref{eq:boltzmann}, the first term describes oscillations in vacuum in the presence of $CP$ violation and decays. The second term accounts for scatterings, where $\Gamma_\pm$ is the scattering rate. This term vanishes for scatterings that are symmetric under $\psi_{\bino}\to \psi_{\bino}^C$. The last term is for annihilations, and $\langle\sigma v\rangle_\pm$ is the thermally-averaged annihilation cross section.

\begin{figure}
    \centering
    \includegraphics[width=\columnwidth]{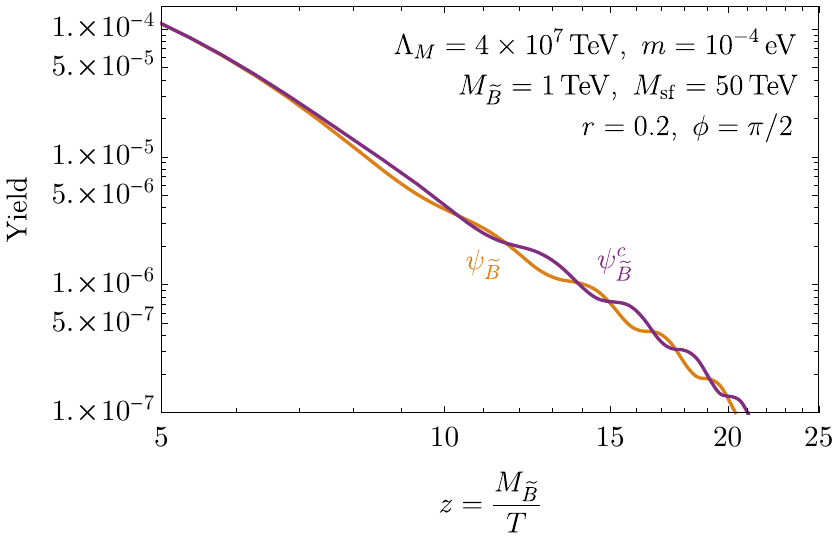}
    \caption{The bino (orange) and antibino (purple) abundances for an initial asymmetry of zero. 
    For $CP$ violating parameter we take $r=0.2$ and $\sin\phi=1$. With $M_{\bino}=1$~TeV and $\Lambda_M=4\times 10^7$~TeV, we have $\Gamma_{\bino}\simeq 2.5\times 10^{-5}$~eV. 
    }
    \label{fig:abundbinoantibino}
\end{figure}

It is illuminating to write the Boltzmann equations in terms of the following redefined densities:
\begin{equation}
\begin{aligned}
    &\Sigma=Y_{\psi_{\bino}}+Y_{\psi_{\bino}^C}\,, ~~~~~~~~~~  \Delta=Y_{\psi_{\bino}}-Y_{\psi_{\bino}^C}\,, \\ 
    &\Pi=Y_{\psi_{\bino}\psi_{\bino}^C}+Y_{\psi_{\bino}^C\psi_{\bino}}\,,~~~~   \Xi=Y_{\psi_{\bino}\psi_{\bino}^C}-Y_{\psi_{\bino}^C\psi_{\bino}}\, .
\end{aligned}
\end{equation}
Using the above, \cref{eq:boltzmann} becomes,
\begin{equation}
\begin{aligned}\label{eqn:BEcoupled}
	zH\frac{d\Sigma}{dz}=& -\Gamma_{\bino}\Sigma + 2 \Gamma_{\bino} r(i\sin\phi\,\Xi -\cos\phi\, \Pi) \\
	&-\frac{s}{2}(\langle\sigma v\rangle_{\rm sens}^{\rm ann}-\langle\sigma v\rangle_{\rm blind}^{\rm ann}) (\Xi^2-\Pi^2)  \\
		&-\frac{s}{2}(\langle\sigma v\rangle_{\rm sens}^{\rm ann}+\langle\sigma v\rangle_{\rm blind}^{\rm ann})(\Sigma^2 -\Delta^2-4Y_{\rm eq}^2)\,,  \\
		zH\frac{d\Delta}{dz} =& -\Gamma_{\bino}\,\Delta +2im\,\Xi\,,  \\
        zH\frac{d\Xi}{dz} =& -\left(\Gamma_{\bino} +2\Gamma_{\rm sens}^{\rm scat}+s\,\Sigma\,\langle\sigma v\rangle_{\rm blind}^{\rm ann}\right)\,\Xi	\\
        &+2im\,\Delta-2i \Gamma_{\bino} r\sin\phi\,\Sigma\,, \\
        zH\frac{d\Pi}{dz} =& -\left(\Gamma_{\bino} +2\Gamma_{\rm sens}^{\rm scat}+s\,\Sigma\,\langle\sigma v\rangle_{\rm blind}^{\rm ann}\right)\,\Pi \\
        & -2\Gamma_{\bino} r\cos\phi\,\Sigma\,.
\end{aligned}
\end{equation}
Note that the scattering rates are defined as
\begin{align}
    \Gamma^{\rm scat} =  \frac{3\zeta(3)}{2\pi^2}\left(\frac{M_{\bino}}{z}\right)^3 \langle\sigma v\rangle^{\rm scat}\,,
\end{align}
where the relevant cross sections are given in \cref{eq:xsections}.
As a representative example, in \cref{fig:abundbinoantibino} we plot the bino and antibino yields for $M_{\bino}=1$~TeV, $M_{\rm sf}=50$~TeV, $\Lambda_M=4\times 10^7$ TeV and $m=10^{-4}$~eV. It can be seen in \cref{fig:abundbinoantibino} that the oscillations start around $z\sim 12$, corresponding to $T\sim 80$~GeV, which is below $T_{\rm sph}\simeq 130$ GeV. This set of parameters was chosen to showcase the oscillations with $CP$ violation. In the next section, we will focus on larger $m$ values so that bino oscillations start before the sphalerons shut off. 
 
Let us estimate when the oscillations commence. Coherent bino--antibino oscillations can only start after the Hubble rate falls below the oscillation frequency, $\omega_{\rm osc}=\Delta m > H(T)$. However, bino interactions with the SM plasma tend to hinder oscillations as well. This phenomenon is an example of the quantum Zeno effect~\cite{10.1063/1.523304}, in which scatterings and annihilations can act as measurements on a quantum system. This effect can be analytically understood by looking at the evolution equations for $\Delta(z)$ and $\Xi(z)$ above. For an approximate analysis, we can omit $\Gamma_{\bino}$ as decays and oscillations are independent. We can then solve for $\Xi$ from the second equation in \cref{eqn:BEcoupled} and use this in the third equation. After a change of variables with $y=z^2$, we get
\begin{align}
    \frac{d^2\Delta}{dy^2}+\frac{\tilde{\Gamma}(y)}{2yH}\frac{d\Delta}{dy}+\frac{m^2}{y^2H^2}\Delta =0\,,
\end{align}
where $\tilde{\Gamma}(y)=2\,\Gamma_{\rm sens}^{\rm scat}+s\,\Sigma\,\langle\sigma v\rangle_{\rm blind}^{\rm ann}$. (Note that the combination $yH$ is a constant.) This equation describes a damped harmonic oscillator in which the damping term drops over time.  Oscillations occur only in an under-damped system. In the existence of interactions, the start of oscillations is when the following under-damping requirement is satisfied,
\begin{align}
        \frac{m^2}{y^2H^2} &>\hspace{-0.05cm} \left(\frac{\tilde{\Gamma}(y)}{4yH}\right)^2 \hspace{-0.1cm} \Rightarrow 
        4m > \left(2\,\Gamma_{\rm sens}^{\rm scat}+s\,\Sigma\,\langle\sigma v\rangle_{\rm blind}^{\rm ann}\right)_{z=z_{\rm osc}}\hspace{-0.15cm}.
\label{eq:mcondition}
\end{align}

\begin{figure}[t]
    \centering
    \includegraphics[width=\columnwidth]{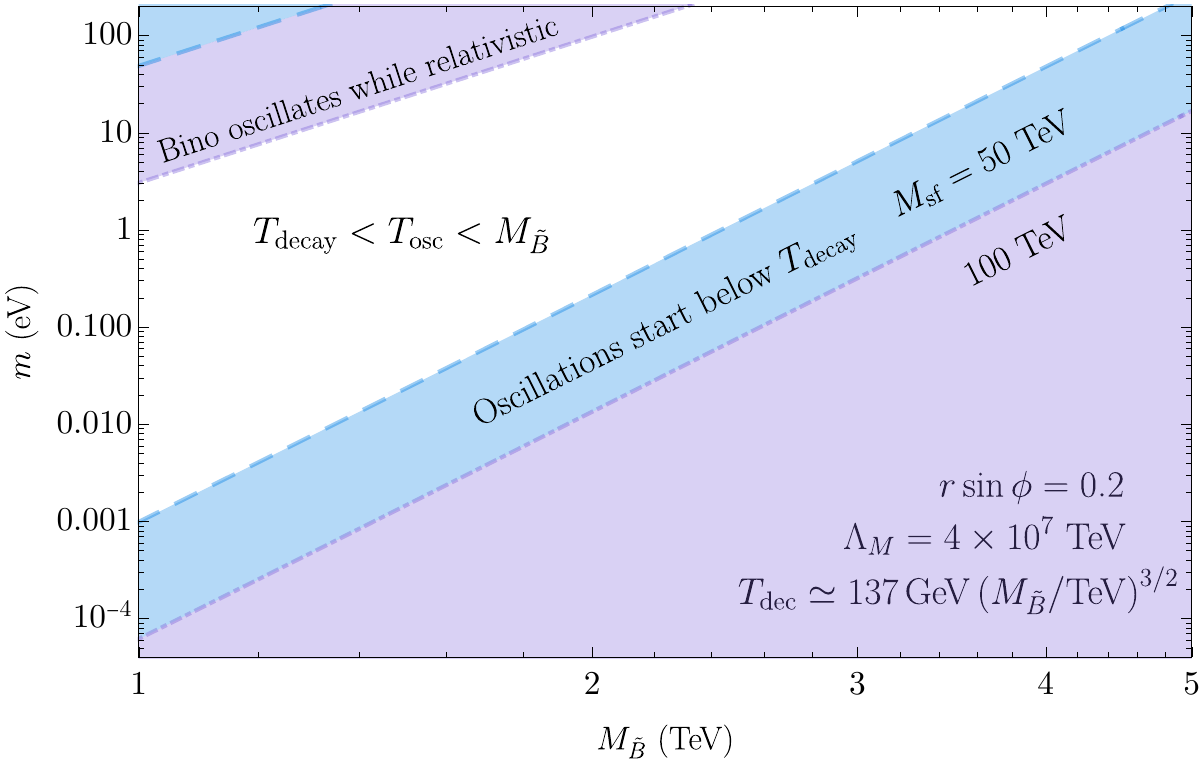}
    \vspace{.1in}
    \includegraphics[width=\columnwidth]{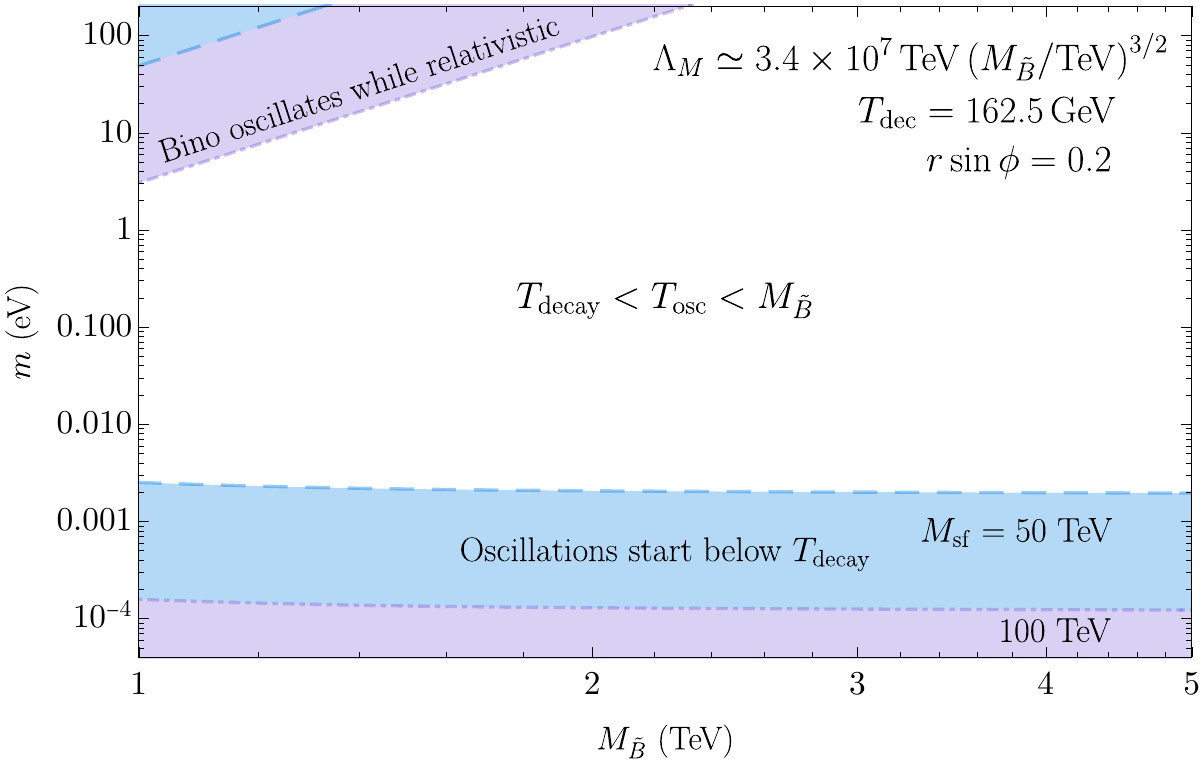}
    \caption{The parameter space, $M_{\bino}$ vs the Majorana mass $m$, for which bino oscillations start between $T=M_{\bino}$ and $T_{\rm dec}$ for $\Lambda_M=4\times 10^7$ TeV with varying $T_{\rm dec}$ (upper) and $T_{\rm dec}\simeq 162.5$~GeV with varying $\Lambda_M$ (lower).  (See \cref{eq:moscregion} and the accompanying text.) Dashed/blue and dashed-dotted/purple lines are for $M_{\rm sf}=50, {\rm ~and~}100$ TeV, respectively. In the bottom-shaded region, bino oscillations start after sphalerons turn off, whereas in the upper-shaded regions, oscillations are expected to start while bino is relativistic, $T>M_{\tilde{B}}$.}
    \label{fig:moscillation}
\end{figure}

If the Majorana mass is large enough, oscillations could start while the bino is relativistic. However, our treatment in \cref{eq:boltzmann} does not capture this regime. To ensure the validity of our analysis, we limit our focus to the parameter space where oscillations begin at temperatures below the bino mass $T_{\rm osc}<M_{\bino}$. Furthermore, it is preferable for the bino to start oscillating before it decays to maximize the \emph{CP} violation. (Since we focus on bino decays that occur before the sphalerons shut off, $T>T_{\rm sph}=130~$GeV~\cite{PhysRevLett.113.141602}, we automatically satisfy $z_{\rm osc}< z_{\rm sph}=M_{\bino}/T_{\rm sph}$.) These requirements translate into a condition on the Majorana mass $m$ range,
\begin{align}
    \tilde{\Gamma}(z=1)> 4m >  \tilde{\Gamma}(z=z_{\rm dec})\,, \label{eq:moscregion}
\end{align}
$z_{\rm dec}$ is when the bino decays. This region is shown in \cref{fig:moscillation}, assuming a thermal bino distribution for the annihilation rate\footnote{For $z=1$, the bino is close to equilibrium and for $z= z_{\rm dec}\gg 1$, scattering rate largely dominates over the annihilation rate. }. In the next section, we will focus on two benchmark scenarios, depicted in the upper and lower plots of \cref{fig:moscillation}. In the first one (upper), we fix $\Lambda_M= 4\times 10^7$~GeV, in which larger bino masses will result in shorter lifetimes. In the second scenario (lower), we fix the bino width such that $T_{\rm dec}\simeq 160$~GeV, which requires larger messenger scales for heavier binos, up to $\mathcal{O}(10^8{\rm~ TeV})$. For the Dirac bino masses we consider $\mathcal{O}({\rm TeV})$, the Majorana masses need to be larger than $\sim 10^{-3}$ eV so that the oscillations are not delayed after the bino starts decaying. In the upper shaded regions, bino is expected to start oscillating while in equilibrium, and our non-relativistic analysis does not apply. (We note that in order to produce the correct baryon asymmetry, we will be far from this near-relativistic region.) The interactions are suppressed with larger $M_{\rm sf}$. Hence, oscillations can start earlier, limiting the region of validity. We will choose $M_{\rm sf}=100$~TeV as a benchmark value in the next section. These conditions could be relaxed with a much more detailed analysis. However, as we show below, a large parameter space where the baryon asymmetry can be generated already exists, even with these limitations.

\section{The Baryon Asymmetry Of the Universe}\label{sec:baryogenesis}

As the binos oscillate and decay in the early universe, the resulting lepton asymmetry can be quantified using the $CP$-violating parameter in \cref{eq:epsilon}\footnote{The asymmetry parameter $\epsilon$ arises from the interference between decays occurring with and without bino–antibino oscillations, as in the neutral $B$-meson system.}:
\begin{align}
     z H(z) \frac{d\Delta_L(z)}{dz}&=\varepsilon\, \Gamma_{\bino}\,\big(\Sigma(z)-2Y_{\rm eq}\big), \label{eq:leptonBoltzman}
\end{align}
where $\Delta_L = Y_L-Y_{\bar{L}}$. (The asymmetry parameter $\epsilon$ is time-integrated, which is a good approximation for the regime we consider where $x\simeq 1-1000$.) Here, we include a washout term from the inverse decays, which is especially important for $M_{\rm sf}\lesssim 50$ TeV, as the bino interactions are large enough to keep them close to equilibrium before they decay. We ignore another washout term coming from $\Delta_L =2$ scatterings, $h \ell \to h\bar{\ell}$ through bino exchange, since these are highly suppressed by a factor of $|M_{\bino}\tilde{m}/\Lambda_M^2|^2$.

The lepton asymmetry is partially converted to a baryon asymmetry through the electroweak sphalerons in the SM. Sphaleron processes conserve $B-L$ while violating $B+L$. By doing a detailed balance of chemical potentials in various electroweak interactions, one finds 
\begin{align}
    \Delta_B=\frac{32+4n_H}{98+13n_H}\Delta_{(B-L)},
\end{align}
where $n_H$ is the number of Higgs doublets. (See, e.g., \cite{Davidson_2008} for a review.) We use a minimally supersymmetric model where $n_H=2$. This conversion ceases when the sphalerons fall out of equilibrium at a temperature of $T_{\rm sph}=130~$GeV~\cite{PhysRevLett.113.141602}. Thus, the resulting baryon asymmetry can be determined as
\begin{align}
    \Delta_B =0.348\times \Delta_L(z_{\rm sph})\,.
\end{align}
\begin{figure}[t]
    \centering
        \includegraphics[width=\columnwidth]{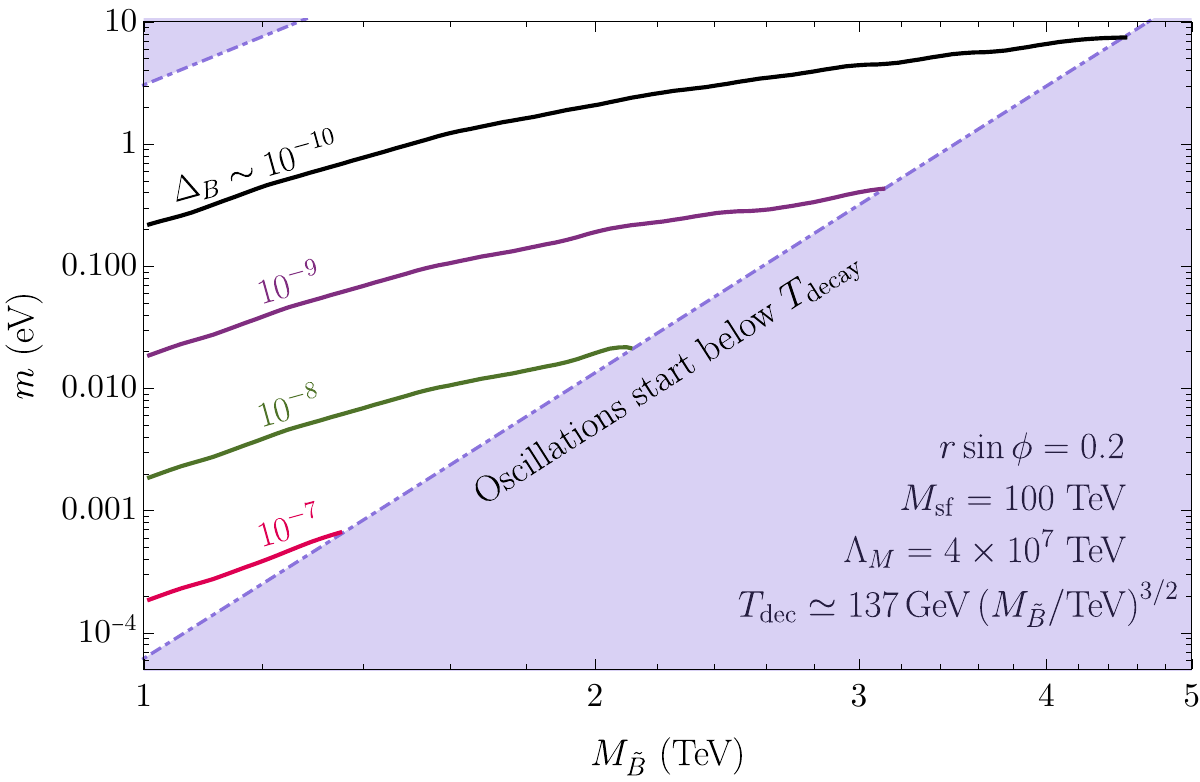}
        \vspace{.1 in}
        \includegraphics[width=\columnwidth]{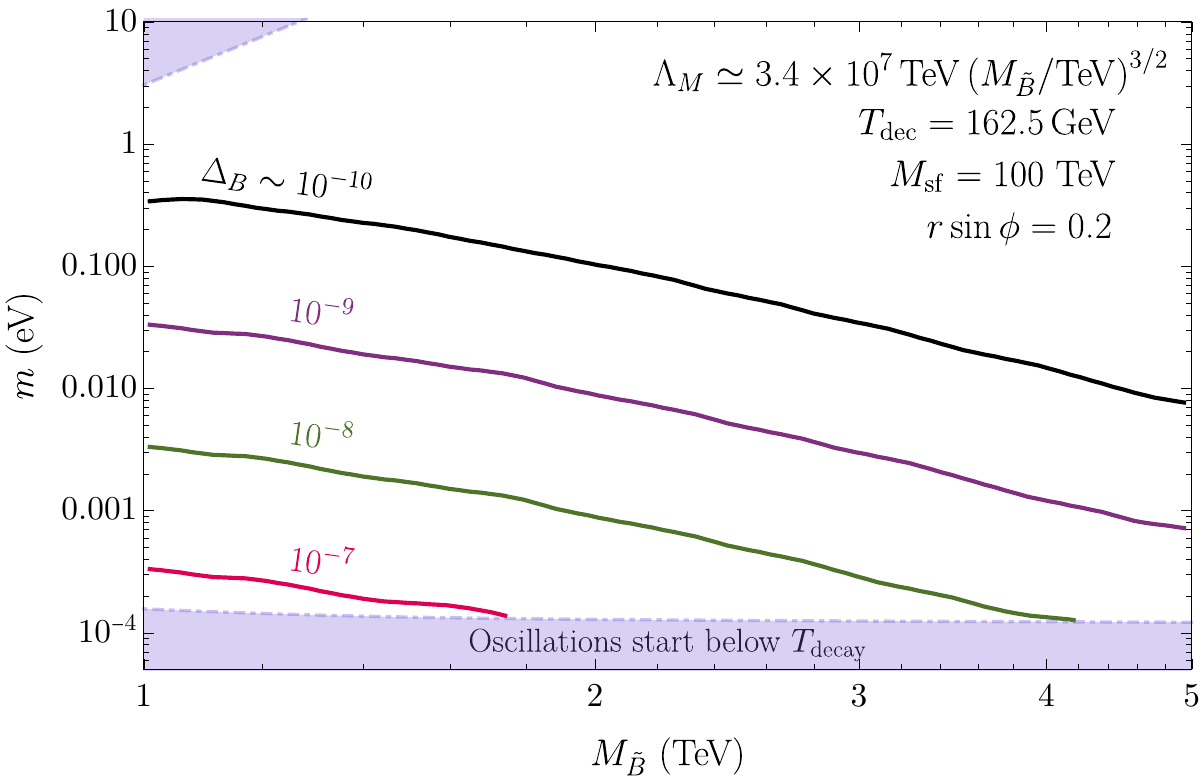}
    \caption{Lines of constant $\Delta_B$ for $\Lambda_M=4\times10^7$~TeV (upper) and $T_{\rm dec}\simeq 162$ GeV (lower) with $M_{\rm sf}=100$~TeV, $r=0.2$ and $\phi=\pi/2$. The observed value is $\sim 8\times 10^{-11}$.}
    \label{fig:basymlines}
\end{figure}
\begin{figure}[t]
    \centering
    \includegraphics[width=\columnwidth]{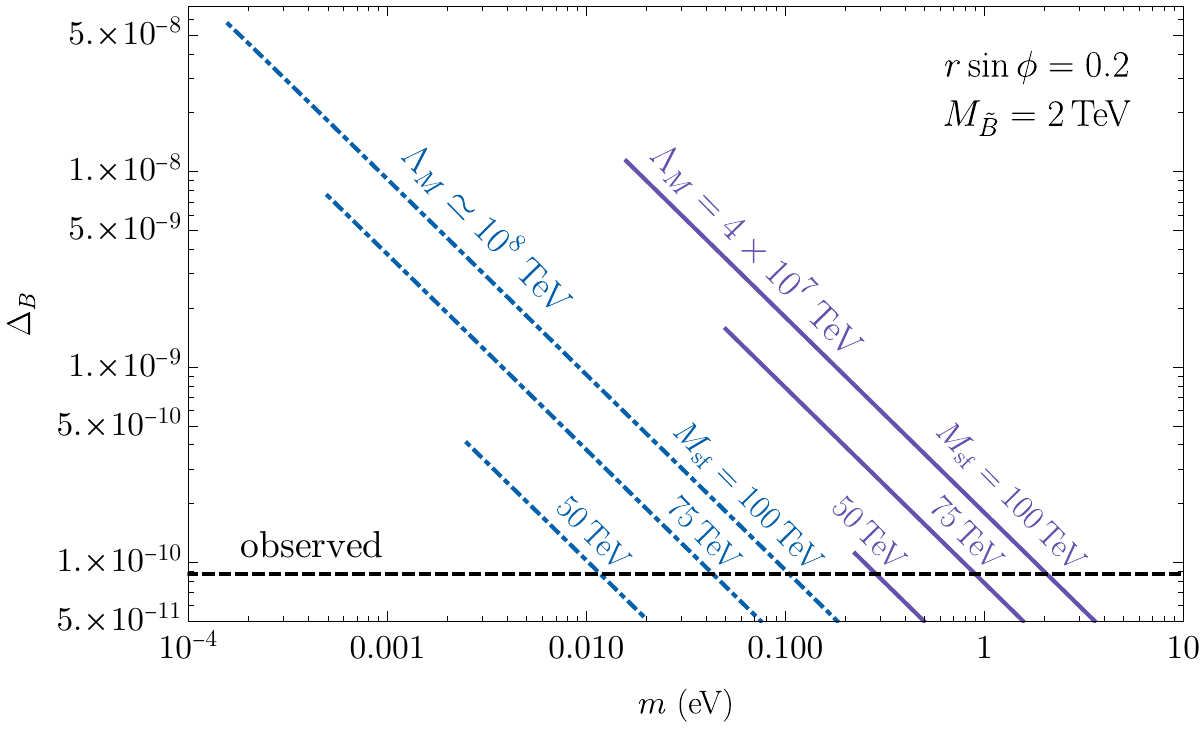}
    \caption{The baryon asymmetry as a function of the Majorana mass 
    $m $ for benchmark parameters $M_{\bino} =2$ TeV and $r\sin\phi=0.2$. The horizontal dashed line represents the observed yield of BAU, $\Delta B\sim 8.7\times 10^{-11}$. Solid/purple lines are for fixed $\Lambda_M=4\times 10^7$ TeV, and dotdashed/blue lines are for fixed bino lifetime $T_{\rm dec}=1.25\, T_{\rm sph}\simeq160 {\rm\, GeV}$, corresponding to $\Lambda_M\simeq 10^8$ TeV for $M_{\rm sf}=50,~75,{~\rm and~}100~ \rm TeV$. The lines are cut off using the condition in \cref{eq:moscregion}.
    } 
    \label{fig:deltaBvsm}
\end{figure}

In \cref{fig:basymlines} we show the lines of constant $\Delta_B$ for varying $M_{\bino}$ and $m$, setting $M_{\rm sf}=100$~TeV, $r=0.2$ and $\sin\phi = 1$ for two scenarios. In the first case, shown in the upper figure, we set $\Lambda_M=4\times 10^7$~TeV and in the second scenario, shown in the lower figure, we fix the bino decays to happen at $T_{\rm dec}\simeq 160$~GeV. In both cases, the baryon asymmetry can exceed the observed value in a significant portion of the parameter space, reaching up to $10^{-7}$, shown as the red lines. For a given value of $M_{\bino}$, the baryon asymmetry increases with decreasing $m$ as $x=2m/\Gamma_{\bino}$ approaches to 1. This allows for smaller values of $r$ and $\phi$, or more specifically, smaller values of the combination $r\,\sin\phi$, which is the relevant combination for $\varepsilon$ (as seen in \cref{eq:epsilon}). For a constant Majorana mass $m$ and constant $\Lambda_M$, heavier binos produce a larger asymmetry because the lifetime $\tau_B \sim \Lambda_M^2/M_{\bino}^3$ is shorter and the bino decays while its number density is larger. Smaller sfermion masses result in a smaller asymmetry due to larger annihilation/pair-production cross-sections, which keeps the bino in equilibrium longer. This feature can be seen in \cref{fig:deltaBvsm} where we plot the baryon asymmetry for $M_{\bino}=2$ TeV and $M_{\rm sf}=50, 75, 100$ TeV. These interactions cease to be relevant for $M_{\rm sf}\gtrsim500~$ TeV.

At this point, let us situate our work in the realm of resonant leptogenesis~\cite{Pilaftsis:2003gt, Pilaftsis:2004xx, Pilaftsis:2005rv, Chauhan:2021xus, Klaric:2021cpi, King:2024idj, Das:2024gua} and ARS leptogenesis~\cite{Akhmedov:1998qx, Drewes:2017zyw, Drewes:2021nqr, Caputo:2018zky, Baumholzer:2018sfb}. Arguably, the most notable difference of our model is the production of heavy neutral leptons, which are binos in our case. Binos can always be produced copiously in the early universe through their non-neutrino interactions with the SM. After their production, their out-of-equilibrium decay is governed by their mixing with neutrinos, as in other leptogenesis scenarios. Hence, unlike the freeze-in scenario of ARS leptogenesis, we are always in the freeze-out regime as in resonant leptogenesis. An additional effect that arises from considering binos as right-handed neutrinos is their scattering off the SM plasma through sfermion exchange. These interactions delay bino--antibino oscillations, which we use to limit our parameter space to the regime where binos oscillate only when non-relativistic. Lastly, in the hybrid inverse seesaw mechanism, the wino interactions are the main source of neutrino masses, while the long-lived bino is responsible for creating the matter--antimatter asymmetry. As discussed in \Cref{sec:model}, this sequestering is assumed to ensure that the bino decays at $T< M_{\bino}$ instead of the strong washout regime. 

\section{Signals}
\label{sec:signals}
The generation of the BAU imposes strong requirements on the parameter space of this model. We have focused on the bino and wino with masses $\mathcal{O}({\rm TeV})$. In order to have a long-lived bino that decays out of equilibrium, we need the bino to be either the lightest neutralino or at least $M_{\bino}<M_{\widetilde{W}}+M_Z$ and a high messenger scale, $\Lambda_M \gtrsim 10^7$ TeV.  In~\cite{Gehrlein:2021hsk}, the LHC phenomenology of a long-lived bino was studied in a $U(1)_{R-L}$-symmetric MSSM model like this one, for messenger scales of up to $10^{11}$~TeV with signals inside the ATLAS detector or at proposed detectors such as MATHUSLA~\cite{Curtin:2018mvb, MATHUSLA:2022sze} and CODEX-b~\cite{Gligorov:2017nwh, Aielli:2022awh}.  However, those results were obtained for a scenario in which binos were produced in squark decays. In this leptogenesis scenario, out-of-equilibrium conditions require heavy sfermions with masses $M_{\rm sf}\gtrsim 50$ TeV, which are inaccessible at the LHC.

For possible collider signatures, a promising mass spectrum is one with decoupled sfermions and gluinos with $\mathcal{O}({\rm TeV})$ neutralinos. This narrows the searches to the wino and the bino, both of which can decay via mixing with the neutrinos. (We take one of the neutralinos to be mostly bino and another one to be mostly wino.) There are two different signal regimes depending on the mass ordering between the bino and the wino.

\begin{enumerate}
    \item $M_{\widetilde{W},\chi^{\pm}}> M_{\tilde{B}}+M_Z$: Winos and charginos can be produced in EW processes and decay to a bino and a gauge boson, e.g. $\widetilde{W}\to Z \widetilde{B}$. (Wino can also decay directly to $Z\nu$ via mixing with neutrinos, but these channels are suppressed by the mixing angles.) Afterwards, the bino decays to the final states of  $W^\pm \ell^\mp, Z\nu, h\nu$. Since $\Lambda_M$ is very large, the bino can be long-lived at collider scales. The expected signal is a prompt vertex accompanied by a displaced vertex.

     \item $M_{\widetilde{W}} +M_Z> M_{\tilde{B}}> M_{\widetilde{W},\chi^{\pm}}$: In this case, the bino,  wino, and charginos will all decay only via their mixing with neutrinos and charged leptons, respectively. For prompt wino/chargino decays, an ATLAS search for trilepton resonances with $139~ {\rm fb}^{-1}$ data excludes wino/chargino with masses between 100 GeV and 1.1 TeV~\cite{ATLAS:2020uer}. Here, in contrast to \cite{Ayber:2023diw}, the bino and wino can both be long-lived due to the large messenger scale. Hence, these constraints might not apply.
\end{enumerate}

\section{Conclusions and Outlook} 
\label{sec:conclusion}

 We studied a $U(1)_{R-L}$-symmetric supersymmetric model that can generate both the light neutrino masses and the baryon asymmetry of the universe. It has been shown that the pseudo-Dirac bino and wino in this model can result in a hybrid type I + III inverse seesaw mechanism that explains the neutrino masses~\cite{Ayber:2023diw}. In this work, we separated the two mechanisms: the wino is responsible for the neutrino mass generation, whilst the pseudo-Dirac bino generates the BAU through its out-of-equilibrium decay in the early universe. The leptogenesis scenario we studied requires a bino of mass a few TeV with a small Majorana mass of $10^{-4}~{\rm eV}-10$ eV. The smallness of the Majorana gaugino masses can be related to how the  $U(1)_{R-L}$ is broken and how this breaking is mediated to the supersymmetric sector.

Pseudo-Dirac fermions can oscillate between their particle and antiparticle states with an oscillation frequency $\omega \sim 2m$ where $m$ is the Majorana mass. If the particle and antiparticle states can decay into the same final state, these oscillations can enhance the $CP$ violation in the system. In this work, we used decays facilitated by bino-neutrino mixing terms in the neutrino mass Lagrangian given in \cref{eq:massLag}. This is a leptogenesis scenario, and the bino needs to decay before EW sphalerons turn off at $T_{\rm sph}\simeq 130$ GeV. Together with the out-of-equilibrium condition, successful leptogenesis requires a messenger scale $\Lambda_M\sim \mathcal{O}(10^8~{\rm TeV})$ for a bino of mass a few TeV. 

The oscillations in the early universe are affected by both the expansion of the universe and interactions between particles and the rest of the plasma. Unlike the most generic leptogenesis scenarios, where the right-handed neutrinos interact with the SM only through their mixing with the light neutrinos, bino has much more effective interactions as it couples to a sfermion and a fermion. These interactions reduce the resulting baryon asymmetry in two ways. They keep the bino in thermal equilibrium, washing away the asymmetry and delaying oscillations, resulting in smaller $CP$ violation. Even with these hindrances, a large baryon asymmetry, $\Delta_B \sim 10^{-7}$, can be produced for $M_{\rm sf}\gtrsim 100$ TeV and $m\sim 10^{-4}$ eV.

Generating the observed BAU imposes stringent constraints on the mass spectrum of the model, pushing it toward a decoupled scenario with very heavy sfermions, $M_{\rm sf}\gtrsim 50$ TeV. Still, the wino can have interesting signatures at the LHC. If the wino is the lightest accessible SUSY particle, it decays to final states with leptons, quarks, and missing energy through its mixing with the SM neutrinos. Since the creation of the BAU requires a high messenger scale, these decays could result in displaced vertices. Furthermore, the branching fractions to different lepton flavors point towards the neutrino mass generation mechanism. These signatures are interesting targets for long-lived particle searches for both the current LHC detectors \cite{ATLAS:2024qoo, ATLAS:2024ocv, CMS:2024qxz, CMS:2024xzb} and for proposed detectors like MATHUSLA \cite{Curtin:2018mvb,MATHUSLA:2022sze} and CODEX-b \cite{Aielli:2022awh}. We leave the collider studies to future work.

\section*{Acknowledgements}
This work was supported by the Natural Sciences and Engineering Research Council (NSERC) of Canada and the  Arthur B. McDonald Canadian Astroparticle Physics Research Institute. SI is grateful for the hospitality of Perimeter Institute, where part of this work was carried out. Research at Perimeter Institute is supported in part by the Government of Canada through the Department of Innovation, Science and Economic Development and by the Province of Ontario through the Ministry of Colleges and Universities. This work was supported by a grant from the Simons Foundation (1034867, Dittrich).

\bibliography{references}

\end{document}